\newcommand{\CLASSINPUTtoptextmargin}{0.8in}
\newcommand{\CLASSINPUTbottomtextmargin}{0.98in}
\documentclass[conference]{IEEEtran}
\IEEEoverridecommandlockouts
\usepackage{cite}
\usepackage{amsmath,amssymb,amsfonts}

\usepackage{color,soul}

\usepackage[ruled,linesnumbered]{algorithm2e}

\usepackage{algorithmic}

\usepackage{graphicx}
\usepackage{textcomp}
\usepackage{xcolor}

\usepackage{mathtools, stmaryrd}
\usepackage{xparse} 
\usepackage{mathtools, nccmath,amsmath,amssymb,amsfonts,amstext, dsfont, color}
\usepackage{subcaption}
\usepackage{float}

\usepackage{array,multirow}

\newtheorem{lemma}{Lemma}

\newtheorem{remark}{Remark}

\newcommand\myon{ \left(\frac{1}{\lambda_n^e}\cdot \frac{\lambda_n^r } {\e^{\lambda_n^r t_C}-1}  + 1 \right)^{-1}}

\newcommand\hit{h}

\newcommand\occupancy{o}
\newcommand\newoccupancy{o}

\newcommand\givenbase[1][]{\:#1\lvert\:}
\let\given\givenbase

\DeclarePairedDelimiterX\Basics[1](){\let\given\sgiven #1}

\newcommand{\Proba}[1]{\mathrm{Pr}\left(#1\right)}
\newcommand{\E}[1]{\mathbb{E}\left[#1\right]}
\newcommand{\Vecnot}[2]{\left(#1\right)_{#2}} 
\newcommand{\Ton}{T^{\mathrm{On}}_{n} }
\newcommand{\Toff}{T_{n}^{\mathrm{Off}} }
\newcommand{\Nopen}[1]{\mathcal{N}(#1)}
\newcommand{\Nclosed}[1]{ \mathcal{N}[#1]}
\newcommand{\Nopenin}[2]{\mathcal{N}_{#1}(#2)}
\newcommand{\Nclosedin}[2]{\mathcal{N}_{#1}[#2]}

\newcommand{\e}{\mathrm{e}}
\newcommand{\exP}[1]{\exp\left( #1\right) }

\DeclarePairedDelimiterX{\Iintv}[1]{\llbracket}{\rrbracket}{\iintvargs{#1}}
\NewDocumentCommand{\iintvargs}{>{\SplitArgument{1}{,}}m}
{\iintvargsaux#1} %
\NewDocumentCommand{\iintvargsaux}{mm} 
{#1\mkern1.5mu..\mkern1.5mu#2}

\DeclareMathOperator*{\argmax}{arg\,max}
\DeclareMathOperator*{\argmin}{arg\,min}

\def\BibTeX{{\rm B\kern-.05em{\sc i\kern-.025em b}\kern-.08em
    T\kern-.1667em\lower.7ex\hbox{E}\kern-.125emX}}
\begin{document}

\title{Computing the Hit Rate of  Similarity Caching} % Policies }
\author{\IEEEauthorblockN{Younes Ben Mazziane\IEEEauthorrefmark{1}, Sara Alouf\IEEEauthorrefmark{1}, Giovanni Neglia\IEEEauthorrefmark{1} and Daniel Sadoc Menasche\IEEEauthorrefmark{2}}
\IEEEauthorblockA{\IEEEauthorrefmark{1}Université C\^{o}te d'Azur, Inria, France, %\\
email: name.surname@inria.fr}
\IEEEauthorblockA{\IEEEauthorrefmark{2}Federal University of  Rio de Janeiro, UFRJ, Brazil, %\\
email: sadoc@dcc.ufrj.br}
}

\maketitle

\begin{abstract}
Similarity caching allows requests for an item 
\(i\)  to be served by a similar item   \(i'\). Applications include recommendation systems,
multimedia retrieval, and machine learning. Recently, many similarity
caching policies have been proposed, but still we do not know how to
compute the hit rate  even for the simplest policies, like SIM-LRU and
RND-LRU that are straightforward modifications of
classical caching algorithms. This paper proposes the first algorithm to
compute the hit rate  of similarity caching policies under the
independent reference model for the request process. In particular, our
work shows how to extend the popular TTL approximation from classic caching to similarity
caching. The algorithm is evaluated on both synthetic and real world traces. 
\end{abstract}


\section{Introduction}
Many applications require to retrieve items similar to a given user's request. For example, in content-based image retrieval~\cite{falchi2008metric} systems, users can submit an image to obtain other visually similar images. A similarity cache may intercept the user's request, perform a local similarity search over the set of locally stored items, and then, if the search result is evaluated satisfactory, provide it to the user. The cache thus may speed up the reply and reduce the load on the server, at the cost of providing items \emph{less similar} than those the server would provide. 
Originally proposed for content-based image retrieval~\cite{falchi2008metric} and contextual advertising~\cite{pandey2009nearest}, similarity caches are now a building block for a large variety machine learning based inference systems for recommendations~\cite{sermpezis18},  image recognition~\cite{drolia2017precog,venugopal2018shadow} and network traffic~\cite{dlcaching22} classification. In these cases, the similarity cache stores past queries and the respective inference results to serve future similar requests.

Motivated by the large number of applications, recently much effort has been devoted to formalize similarity caching~\cite{neglia2021similarity,garetto2021content} as well as to propose new caching policies~\cite{zhou2020adaptive,sabnis2021grades,salem2021accai}. Despite this research, to the best of our knowledge, we still do not know how to compute basic performance metrics---like the percentage of requests satisfied by the cache---even for the simplest similarity caching policies, like SIM-LRU and
RND-LRU, which were proposed in the seminal paper~\cite{pandey2009nearest}. SIM-LRU and
RND-LRU are variants of the basic LRU policy, but are much more challenging to analyze than LRU due to strong coupling across items in the cache. In fact, in classic caching, an item in the cache only contributes to serve requests for the very same item, while in similarity caching, the same item can serve requests for a set of similar items as far as neither them, nor their most similar items, are stored in the cache. It follows that, in similarity caching, the number of requests satisfied by an item in the cache depends in general on \emph{the whole cache state}.

In this paper, we introduce the first algorithm to estimate SIM-LRU and
RND-LRU hit rate  under the independent reference model (IRM) \cite{irm-fagin-1977} for the request process. The algorithm  alternates between two steps. In the first step, given a tentative estimate of the rate of requests served by each item when present in the cache, the occupancy probability  of each item (i.e., the probability that the item is in the cache) is computed relying on the well-known TTL approximation~\cite{irm-fagin-1977,che2002hierarchical} (also known as Che's approximation), which has been successfully used to study classic caching policies. In the second step, the current vector of occupancy probabilities, and similarity relations across items, are used to update the rate of requests served by each item. Our experiments both on synthetic traces and a realistic trace for a recommendation system show that our algorithm provide accurate estimates of the hit rate, definitely more precise than other intuitive approaches one could think about.

The paper is organized as follows:   background and notation are introduced in Sec.~\ref{s:background-notation}, our algorithm for computing the hit rate for SIM-LRU and RND-LRU is presented in Sec.~\ref{s:TTL-SIM-LRU},  its performance is evaluated on both synthetic and real word traces in Sec.~\ref{s:experiments} and Sec.~\ref{Conclusion} concludes. 
%We consider a cache deploying similarity caching: whenever a request for an item is issued, it may be served by a reply from a similar item.  We leverage Che's approximation to compute item occupancies:
%\begin{itemize}
%    \item effective arrival rate towards each item is fixed and given:  compute occupancies assuming that metric space between items (routing matrix between items) is fixed and given. In addition, the fraction of requests for an item that is served by another item is known  
%    \item recompute effective arrival rate towards each cached item:  recompute the fraction of requests issued to an item but served by another  item 
%\end{itemize}
%Note that the metric space between items  is assumed to be  fixed and given

%As recognized in \cite{garetto2020similarity}, the idea of similarity caching has been rediscovered a number of times under different names:  recognition caches~\cite{drolia2017cachier, drolia2017precog}, approximate deduplication~\cite{guo2018potluck}, semantic caches~\cite{venugopal2018shadow}, prediction caches~\cite{crankshaw17}, approximate caches~\cite{kumar2019accelerating}, and soft caches~\cite{sermpezis18,spyropoulos16}.

\begin{table}[t]
  \caption{Table of notation} \centering
\begin{tabular}{ |l|l| } 
 \hline
 Variable & Description \\
 \hline
 %\hline
 \multicolumn{2}{c}{{\bf Basic parameters}} \\
 \hline
 $I$ & set of items \\
 $N=|I|$ & catalogue size \\
 $C$ & cache capacity \\
 $S$ & state of cache; set of cached items \\
  $\lambda_n$ & arrival rate of requests to item $n$ \\
 \hline
 \multicolumn{2}{c}{{\bf Similarity cache parameters}} \\
\hline
 $\mathrm{dis}(\cdot, \cdot)$ &function measuring the dissimilarity between items \\
 $d$ & threshold similarity \\
  $\Nopen{n}$ & neighbours of item $n$ \\ 
 $\Nclosed{n}$&  neighbours of item $n$ including $n$ \\
 $\Nopenin{i}{n}$ & items in $\Nopen{n}$  strictly closer to $n$ than $i$ \\
 $\Nclosedin{i}{n}$ & items in $\Nclosed{n}$ strictly closer to $n$ than $i$ \\
 %strictly closer items to $n$ than $i$ among items in $\Nclosed{n}$\\
%  $c_{S}(n)$& closest item to $n$, stored in the cache\\
 $q_n(i)$& probability of an approximate hit for $i$ by $n$\\
 \hline
  \multicolumn{2}{c}{{\bf Inferred variables}} \\ \hline
  $ \lambda^e_n$ & insertion rate of  item $n$; rate at which it enters the cache \\
  $\lambda^r_n$ & refresh rate of item $n$ \\
%   $\lambda=\sum \lambda_n$ & Total arrival rate \\
%  $p_n=\lambda_n/\lambda$& Probability that an item is requested \\
%   $\lambda^e_n$ & Entry rate of item $n$ \\  
%   $\lambda^r_n$ & refresh rate of item $n$ \\
  $t_C$ & characteristic time \\
  \hline
  \multicolumn{2}{c}{{\bf Key metrics of interest}} \\
\hline
$h_n$ & probability of an approximate hit of a request for item $n$ \\
%& (either $n$ or one of  its neighbors is cached)\\ 
$o_n$ & fraction of time  that item $n$ is     cached  \\
$H$ & cache hit probability \\
 \hline
\end{tabular}  
\end{table}

\section{Background, Notation and Assumptions}
\label{s:background-notation}
\subsection{Similarity Caching}
 
 %A \textbf{similarity search} consists in answering a query
 In \textbf{similarity search} systems users can request to a 
 remote server, storing a set of items $I$, the $k$ most  similar items to a given item $n$, given a specific definition of similarity.
 %consists in requesting a answering a query for an item $n$ from a remote server containing a set of items $I$, by the most $k$ similar items in the server given a specific definition of similarity. 
 In practice items are often represented by vectors in $\mathbb{R}^{d}$ (called embeddings) \cite{mcauley2015image} so that the dissimilarity cost, $\mathrm{dis(.,.)}: I^2 \xrightarrow[]{} \mathbb{R}^{+}$, can be selected to be an opportune distance between the embeddings, e.g., the Euclidean one. A cache, that stores a small fraction of the catalog~$I$, could be deployed next to the users to reduce the fetching cost of similarity searches.
The seminal papers \cite{falchi2008metric,pandey2009nearest} suggest the cache may answer a request using a local subset of items potentially different from the true closest neighbors to further reduce the fetching cost while still maintaining an acceptable dissimilarity cost. They refer to such caches as \textbf{similarity caches}.

One of the popular dynamic similarity caching policies is SIM-LRU \cite{pandey2009nearest}. This policy maintains an ordered list of $C$  key-value pairs. Each key is the embedding of an item $n$ requested in the past and its corresponding value is a list containing the $k'\geq k$ closest items to $n$ in $I$. We denote by $S$ the set of keys stored in the cache.
Upon a similarity search for an item~$n$, SIM-LRU selects the closest local key to $n$, i.e., $\hat n \triangleq \argmin_{m\in S} \mathrm{dis}(n,m)$. If the dissimilarity cost between $n$ and $\hat n$ is smaller than a threshold $d>0$ ($\mathrm{dis}(n,\hat n)\leq d$), the request experiences an \textit{approximate hit}:\footnote{
    Note that we have an exact hit if $\hat n= n$.
}
the cache replies to the request for $n$ selecting the $k$ closest items to $n$ among the $k'$ values stored for $\hat n$ and moves $\hat n$'s key-value pair to the front of the list. Otherwise, the request experiences a \emph{miss}: it is forwarded to the original server to retrieve the $k'$ closest items to $n$, out of which the closest $k$ are provided to the user. The cache then adds the new key-value pair for $n$ to the front of the list and evicts the key-value pair at the bottom of the list. %  the closet item stored in the cache $n$ to $o$ is used to answer the request for $o$ if their dissimilarity cost is smaller than a threshold $d$, i.e. $\mathrm{dis}(n,o)\leq d$. In this case an \textit{approximate} hit happens and the similarity search is answered by $n$. Otherwise it is a miss, and the request is forwarded to the original server. The ordered list is updated on an LRU fashion for an approximate hit or a miss.
We observe how the use of key-value pairs in SIM-LRU essentially converts the search of $k$ closest items into the search of the closest key in the cache. For simplicity's sake, from now on we will just identify the items, their keys and the corresponding values and say for example that the cache replies to a request for $n$ with the closest item $\hat n$ in the cache.

RND-LRU \cite{pandey2009nearest} is a  generalization of SIM-LRU, where $\hat n$ is used to reply to a query for $n$ with a probability $q_{\hat n}(n)$ which decreases with their dissimilarity, and it is in any case null for $\mathrm{dis}(\hat n,n)> d$. We retrieve the behaviour of SIM-LRU when $q_{\hat n}(n)=1$ if $\mathrm{dis}(\hat n ,n)\leq d$.

%  decides an approximate hit for an item $i$ by one of its cached neighbours $n\neq i$,  with a probability $q_{n}(i)$, that is decreasing with the dissimilarity between the requested item and the closest item in the cache.

\subsection{Our Assumptions}
We assume that requests follow a Poisson process with request rate $\lambda_n$ for item $n$, and each request is independent from the previous ones, i.e., requests follow the Independent Reference Model (IRM)~\cite{irm-fagin-1977}.
Under SIM-LRU or RND-LRU, a request for item $n$ could be served by any item closer than $d$ to $n$. We denote the set of such items  as $\Nclosed{n}\triangleq \{m\in I: \; \mathrm{dis}(n,m) \leq d\}$. We call the elements in $\Nclosed{n}$ distinct from $n$ the \emph{neighbours} of $n$ and we denote their set as $\Nopen{n}\triangleq \Nclosed{n} \setminus \{n\}$. For the sake of simplicity, we assume that items in $\Nopen{n}$ can be strictly ordered according to their dissimilarity wrt $n$, i.e., for any  $(i,j)\in \Nopen{n}$ and $i\neq j$, we have $\mathrm{dis}(n,i)\neq \mathrm{dis}(n,j)$. If this is not the case, we can introduce an arbitrarily order for items with the same dissimilarity. For convenience, we also define in a similar way the sets $\Nopenin{i}{n}$ and $\Nclosedin{i}{n}$, subsets of $\Nopen{n}$ and $\Nclosed{n}$, resp., designating items that are closer to $n$ than~$i$, i.e. $\Nopenin{i}{n}\triangleq \{ m\in \Nopen{n}: \; \mathrm{dis}(n,m)< \mathrm{dis}(n,i) \}$ and $\Nclosedin{i}{n}\triangleq \{ m\in \Nclosed{n}: \; \mathrm{dis}(n,m)< \mathrm{dis}(n,i) \} $.

\subsection{TTL Approximation for LRU Cache}
\label{ss:ttl-lru}
% Computing the hit rate of an LRU cache under the IRM model can be done by solving the Markov chain representing the states of the cache at each time step. The problem with this approach is that the space state grow exponentially fast with the cache capacity $C$, making the computation costly.

% References  proposed to estimate 

The hit rate of an LRU cache under the IRM model can be estimated using what is referred to in the literature as the TTL approximation~\cite{irm-fagin-1977,che2002hierarchical}. The approximation considers that any cached item $n$, if not requested, will stay in an LRU cache with capacity $C$ for a time $t_C$ that is deterministic and independent of $n$; $t_C$ is called the cache  `characteristic time'. This approximation has been later supported by theoretical arguments in \cite{fricker2012versatile,jiang18}.
%: Fricker, Robert and Roberts stated that under Zipfian distributions $t_C$'s correlation coefficient converges to $0$ as the cache capacity $C$ goes to infinity. 
Under the TTL approximation,
%and for a Poisson arrivals process, 
a hit occurs for an item if the inter arrival time between two requests for the same item is smaller than $t_{C}$. Thus, the hit probability   $h_n$ can be approximated as: 

\begin{equation}
 \label{e:hit-rate-item-lru-che}
        h_n  \approx 1- \e^{-\lambda_n t_C}~. 
\end{equation}
Since the flow of arrivals is Poisson,  the ``Poisson Arrivals See Time Averages'' (PASTA) property implies that the probability~$\occupancy_n$ that an item $n$ is in the cache (i.e., the occupancy probability, or simply occupancy) is equal to the probability that a request for that same item experiences a hit, i.e. $\hit_n = \newoccupancy_n$. The cache capacity constraint is given  in expectation by the following equality: 
       \begin{equation}\label{e:cache-capacity-constraint}
            \sum_{n\in I}  \newoccupancy_n  = C ~,
        \end{equation}
The above expression allows  us to deduce $t_C$, e.g., by means of a bisection method. The hit rate $H$ can be simply computed as $H=\sum_{n} \lambda_n h_n$ with $h_n$ computed as in \eqref{e:hit-rate-item-lru-che}.

% \subsection{Assumptions}
% \label{ss:assumptions}
% We assume that the arrivals process for each item $n\in I$ is Poisson. For the simplicity of the analysis we further suppose that $w_{n,m}\neq  w_{n,l}, \forall n\neq m \neq l$, i.e. an item $n$ can always choose a better neighbour. Similarly to Che's approximation for regular caching, we assume that the time an item stays in the cache if it is not serving any requests is constant and independent of $n$ and we denote it as $t_C$. Let $S(t)$ be the state of the cache at time $t$ and $\Omega=\{S_1, \ldots, S_{|\Omega|} \}$ the set of all possible states when the cache is managed by SIM-LRU, we assume that there exist a stationary setting, i.e. $\lim_{t\to +\infty} \Proba{S(t)=S_i} = \pi_{i}$ $\forall i \in [\Omega]$, and we have $\sum_{i\in  [\Omega]} \pi_{i} = 1$.     

\section{TTL Approximation for Similarity Caching}
\label{s:TTL-SIM-LRU}
Analogously to LRU, TTL approximation for RND-LRU assumes that the time an item stays in the cache if it is not serving any requests is deterministic and independent of $n$ and we denote it as $t_C$. The hit rate for an item $n$ (i.e.,  the rate of requests incurring an approximate hit) can no longer be computed as in \eqref{e:hit-rate-item-lru-che} as the request for $n$ can be satisfied even if $n$ is not in the cache.
% For SIM-LRU, an approximate hit happens if at least one  of item $n$'s neighbours is in the cache at the moment of the request for item $n$.
Let $S$ denote the current state of the cache, i.e., the set of items it stores.
For RND-LRU, an approximate hit  for item $n$ occurs if at least one of the items in $\Nclosed{n}$ is present in the cache. When the closest item to $n$ present in the cache is $i\in \Nclosed{n}$, i.e. $S\cap \Nopenin{i}{n}=\emptyset$ and $i\in S$,  $i$  serves the request for $n$ with probability $q_i(n)$. Taking advantage of the PASTA property, it follows that $h_n$ for RND-LRU can be expressed as: 
% \begin{equation}\label{e:hit-rate-SIM-LRU}
%     \hit_n = \Proba{ S\cap \Nclosed{n}  \neq \emptyset},
% \end{equation}
\begin{equation}\label{e:hit-rate-SIM-LRU}
    \hit_n = \sum_{i\in \Nclosed{n}} q_{i}(n) \cdot \Proba{ S\cap \Nopenin{i}{n}  = \emptyset, i\in S}.
\end{equation}
% and $\Nopenin{i}{n}$ is a subset of $\Nopen{n}$ containing items strictly closer to $n$ than $i$, more specifically: 
% \begin{align}
%     \label{e:setNinopen}
%     &\Nopenin{i}{n}\triangleq \{ m\in \Nopen{n}: \; \mathrm{dis}(n,m)< \mathrm{dis}(n,i) \}~.
% \end{align}
In what follows, we provide an alternative approach to compute the occupancies for an LRU cache, under the TTL approximation,
% without using the PASTA property 
that is complementary to that presented in Sec.~\ref{ss:ttl-lru}. Then, we identify the differences between RND-LRU and LRU and compute its  occupancies in a similar way.
% to identify the differences between RND-LRU and LRU and compute the occupancies for these policies in a similar way.         

% We first start by writing a system of equations expressing the relation between the occupancies, we propose to solve this system of equations using fixed point method, then we deduce $h_n$ and the final hit rate.   

\subsection{Occupancies and Hit Rates} 
To derive the occupancy of an item $n$, we first observe that the instants when item $n$ is evicted from the cache are regeneration points for a renewal process \cite{resnick2007heavy}. A renewal cycle consists of two consecutive time periods: a time period of duration $\Toff$, that starts immediately after item $n$ is evicted from the cache and ends when it re-enters the cache, and  a time period of duration $\Ton$, that ends when item $n$ is evicted again from the cache.
%During this cycle, we denote the time spent inside the cache as $\Ton$ and the time spent outside the cache as $\Toff$. 
From the renewal theorem, the occupancy can be computed as: 
\begin{equation}
\label{e:occupancy}
     \newoccupancy_n = \frac{\E{\Ton}}{\E{\Toff} + \E{\Ton}}~. 
\end{equation}

\paragraph{Expectation of $\Toff$}
% For LRU, $\Toff$ is the waiting time for a request for item $n$ after just being evicted.
% Under the IRM model and thanks to the memoriless property of the exponential distribution, this residual interarrival time is exponentially distributed with rate $\lambda_{n}$, implying that $\E{\Toff} = 1/ \lambda_{n}$.

$\Toff$ is the waiting time for a miss for $n$ after $n$ has been evicted from the cache. For LRU,   under the IRM model,   thanks to the memoryless property of the exponential distribution the residual interarrival time  $\Toff$ is exponentially distributed with rate $\lambda_{n}$, implying that $\E{\Toff} = 1/ \lambda_{n}$.

% For SIM-LRU \cite{pandey2009nearest}, $\Toff$ is the waiting time for a request for item $n$ such that none of $n$'s neighbours is cached at the moment of the request (otherwise, a cached neighbour serves this request and item $n$ is not admitted into the cache).
For RND-LRU, when a request for $n$ finds none of item $n$'s neighbours in the cache, a miss occurs with probability $1$. If instead a request for $n$ finds $i$ to be the closest neighbour of $n$ in the cache, i.e. $\Nopenin{i}{n}\cap S=\emptyset, i\in S$, the probability of a miss is  $1-q_{i}(n)$.   
% even if one of $n$'s neighbours is present in the cache, item $n$ can still be admitted to the cache with probability, where $i$ is the closest neighbour to $n$ present at the cache at the moment of a request for $n$.
Let $p_n^e(i)$ be the probability that $i$ is the closest neighbour to $n$ in the cache, and that a miss occurs, namely: 
\begin{align}
    \label{e:pnei-rnd-def}
    &p_{n}^{e}(i) \triangleq \left( 1-q_{n}(i) \right) \Proba{S \cap \Nopenin{i}{n} =\emptyset, i\in S\given n\notin S}.
\end{align}
The probability $p_n^e$ of a miss for $n$ is then:
% enters the cache when requested
%: $p_n^e$, namely:  
%$\Toff$ is a random sum $F$ of i.i.d.\ random variables $(X_i)_{i\in [F]}$ measuring the interarrival times between item $n$'s requests such that the cache, in all the instants $t_j= \sum_{i=1}^{j} X_i$ where $j\in [F-1]$, does not contain any of item $n$'s neighbours, i.e. $\Nopen{n}\cap S =\emptyset$. We can then estimate the expectation of $\Toff$ as:

% Let $p_n^e$ be the probability that, namely:  
\begin{align} \label{e:pne-def-rnd}
    &p_n^e  \triangleq \Proba{S\cap \Nopen{n}= \emptyset \given n\notin S} + \sum_{i\in \Nopen{n}} p_{n}^{e}(i) ~.
\end{align}
% If item $n$ is requested, $p_n^e(i)$ is the probability that item $n$ is admitted in the cache and $i$ being the closest neighbour of $n$ present in the cache.  
% Let $p_{n}^{e}$ be the probability that none of $n$'s neighbours is cached given that item $n$ is not in the cache, namely, 
% \begin{align} 
% \label{e:pne-def-sim}
%     &p_{n}^{e} \triangleq \Proba{S\cap \Nopen{n}= \emptyset \given n\notin S} ,
% \end{align}
Consequently, when item $n$ is not cached,
% its requests process is thinned with probability $p_{n}^{e}$ and
the rate at which item $n$ re-enters the cache is: 
\begin{align} 
\label{e:lambda-entry-def}
    &\lambda_{n}^{e} = \lambda_n  p_{n}^{e} ~.
\end{align}
Similarly to the case of LRU, we write $\E{\Toff}=1/\lambda_{n}^{e}$.

\paragraph{Expectation of $\Ton$}
For LRU, the time spent by an item in the cache is at least $t_C$, under TTL approximation. Each time the interarrival time between requests for item $n$ is smaller than $t_C$, there is a hit, and $n$ is moved to the top of the queue. In this case,   item $n$ is refreshed, i.e.,  its timer is re-initialized. On the other hand, when this interarrival time is larger than $t_C$, $n$ is evicted from the cache. Time interval  $\Ton$ is the sum of a random number $F$ of time intervals with duration shorter than $t_C$ (corresponding to $F$ consecutive hits) and $t_C$ (the time between the last hit and item's eviction). It follows that
\begin{equation}
\label{e:Ton-sum-value}
    \Ton\approx\sum_{j\in [1..F]} X_j + t_C~,
\end{equation} 
where $(X_j)_{j\in [1..F]}$ are the interarrival times between requests for item $n$ such that $X_j<t_C$  for all $j\in [1..F]$. Since we have: 
\begin{align}
   & \E{X_j\given X_j< t_C} = \frac{1}{\lambda_{n}} - \frac{t_C}{\exp(\lambda_{n} t_C) - 1}  ~,\\
   & \E{F} = \exp(\lambda_{n} t_C)- 1~,
\end{align}
we conclude from Wald's identity and~\eqref{e:Ton-sum-value} that:
\begin{align}
\label{e:Ton-LRU}
    \E{\Ton} \approx  \frac{\e^{\lambda_{n} t_C}-1}{\lambda_{n}}~.
\end{align}
%It is possible to provide a different interpretation for \eqref{e:Ton-LRU}.  Recall that the busy period of a queue corresponds to the contiguous period of time that starts with the arrival of a client to an empty system and ends when the system becomes empty again.  
%Then,  the above expression corresponds to the busy period of an M/G/$\infty$ queue wherein arrival rates and mean service times equal $\lambda_n$ and $t_C$, respectively. Indeed, as soon as a request occurs while the item is stored in cache, the period of time that the item remains in cache is further augmented by $t_C$. The item is evicted when no request to it occurs during an interval of $t_C$ time units.

% For SIM-LRU, item $n$ can be refreshed thanks to its neighbours $i\in \Nclosed{n}$. 
% %if the request for $m$ was received at a time where $n$ was the closest to item $m$ among all its neighbours at the cache, i.e. the request was received when the cache state verified: 
% More specifically, if a neighbour $i$ is requested and cached item $n$ is the closest neighbour to $i$ among all cached items, then item $n$ $(1)$ serves the request for $i$ and $(2)$ is refreshed in the cache. In such a case, the state of the cache verifies $S\cap \Nclosedin{n}{i}=\emptyset$, where:

For RND-LRU, item $n$ is not only refreshed by its own requests but also by requests for its neighbours.
% can be refreshed thanks to its neighbours $i\in \Nclosed{n}$. 
%if the request for $m$ was received at a time where $n$ was the closest to item $m$ among all its neighbours at the cache, i.e. the request was received when the cache state verified: 
More specifically, if item $i\in \Nclosed{n}$ is requested and item $n$ is the closest neighbour to $i$ among all cached items, then item $n$ first  $1)$ serves the request for $i$ with probability $q_{n}(i)$ and then $2)$ is refreshed in the cache. In such a case, the state of the cache verifies $S\cap \Nclosedin{n}{i}=\emptyset$.
% \begin{equation}
% \Nclosedin{n}{i} \triangleq \mathcal{N}_{n}(i) \cup \{i\}~.
%     % \Nclosedin{n}{i} \triangleq \{ m\in \Nclosed{i}: \; \mathrm{dis}(i,m)<\mathrm{dis}(i,n) \}~.
% \end{equation}
% For RND-LRU, even if item $i\in \Nopen{n}$ is requested and $n$ is the closest item present in the cache, $n$ is only refreshed with probability $q_{n}(i)$.
$\Ton$ can still be written as in \eqref{e:Ton-sum-value}. However, the refresh rate for   random variables $X_j$ is no longer $\lambda_n$ but a higher rate $\lambda_{n}^{r}$ expressed as:         

\begin{align}
    \label{e:lambda-refreshement-def}
    &\lambda_{n}^{r} = \sum_{i\in \Nclosed{n}} p_{n}^{r}(i) \lambda_{i}, \\ 
    \label{e:pnr-def}
    & p_{n}^{r}(i) = q_{n}(i) \cdot \Proba{S\cap \Nclosedin{n}{i} = \emptyset \given n \in S} ~. 
\end{align}
(Notice that $\mathcal{N}_{n}[n]=\emptyset$ and then $p_{n}^{r}(n)=1$.)
Similarly to \eqref{e:Ton-LRU}, the expected value of $\Ton$ can be computed as
\begin{equation}
    \label{e:ton-expectation}
    \E{T_n^{\mathrm{On}}} \approx  \frac{\e^{\lambda_{n}^{r} t_C}-1}{\lambda_{n}^{r}}~.
\end{equation}

%\begin{remark}
% For RND-LRU \cite{pandey2009nearest}, a cached item $n$ that serves a request for item $i$ (for which it is the closest neighbour in cache) is refreshed only with probability $q_n(i)$ depending on the similarity between $i$ and $n$. (For SIM-LRU, $q_n(i)=1$.)
%the probability that item $i$ refreshes item $n$, even if $n$ is its closet neighbours in the cache, is not $1$ like SIM-LRU, but depends on the similarity between $i$ and $n$ say $q_n(i)$. 
% Formula \eqref{e:ton-expectation} still holds for RND-LRU but the refresh probabilities $p_n^{r}(i)$ in \eqref{e:pnr-def} are multiplied by $q_n(i)$ : 
%     \begin{equation}
%     \label{e:pnr-RND-LRU}
%         p_n^{r}(i) = q_n(i) \cdot \Proba{S\cap \Nclosedin{n}{i} = \emptyset \given n \in S} ~.
%     \end{equation}
%\end{remark} 

\paragraph{Computing the occupancy}
Replacing the expressions for $\E{\Toff}$ and $\E{\Ton}$ in \eqref{e:occupancy}, we derive the occupancy. For LRU, we obtain the knonw result in~\eqref{e:hit-rate-item-lru-che} (remember that $h_n=o_n$ for LRU).
%$\newoccupancy_n \approx 1-\e^{-\lambda_n t_C}$; see \eqref{e:hit-rate-item-lru-che}. 
For RND-LRU, 
%the occupancy can be deduced from \eqref{e:occupancy} using $\E{\Toff}=1/\lambda_{n}^{e}$ and \eqref{e:ton-expectation}. 
we obtain
\begin{align}
\label{e:occupancy-simlru-che}
    \newoccupancy_n \approx   %\frac{ \E{T_n^{\mathrm{On}}}}{ \E{T_n^{\mathrm{On}}}+1/\lambda_n^e} = 
    \myon.
\end{align}

%\subsubsection{Model 1 : Independent events}
\paragraph{Computing $p_{n}^{e}$, $p_{n}^{r}(i)$ and the hit rate}
Under the classic TTL approximation for LRU, items  are coupled only through the value of the characteristic time $t_C$. Conditioned on~$t_C$,  events related to the presence of items in the cache are independent, e.g.,  $\Proba{n,m \in S} = \Proba{n \in S} \cdot \Proba{m \in S}$. We maintain this independence also for RND-LRU. Then, 
%Should the events $\left(\{u \in S\}\right)_{u\in \bigcup_{j\in \Nclosed{m} }\Nclosed{j}}$ be independent for every item $m$, computing the probabilities $\left(p_{n}^{e}\right)_{n\in I }$ and $\left(p_{n}^{e}(i), p_{n}^{r}(i)\right)_{(i,n) \in I^{2}}$ as in 
\eqref{e:pne-def-rnd},   \eqref{e:pnei-rnd-def}, and \eqref{e:pnr-def} can be written as follows: 
\begin{align}\label{pnei-independence}
& p_{n}^{e}(i) = (1-q_i(n))\cdot \newoccupancy_{i} \prod_{m \in \Nopenin{i}{n}} \left(1-\newoccupancy_{m}\right)~,     \\ \label{e:pne-independence}
 &   p_{n}^{e} =\prod_{m \in \Nopen{n} } (1-\newoccupancy_m) + \sum_{i\in \Nopen{n}} p_n^e(i)~, \\ \label{e:pnr-independence}
 &   p_{n}^{r}(i) =  q_n(i)\prod_{m \in \Nclosedin{n}{i}} \left(1-\newoccupancy_{m}\right)~.
\end{align}
Under the independence assumption,  RND-LRU's hit rate $h_n$ for item $n$'s requests in \eqref{e:hit-rate-SIM-LRU} can be computed as: 
%   \begin{equation}
%       \hit_{n} =  1- \prod_{i\in \Nclosed{n}} \left( 1-\newoccupancy_{n}  \right)~.
%   \end{equation}
\begin{equation}\label{e:hit-rate-rnd-lru-independence}
   \hit_{n} =  \sum_{i\in \Nclosed{n}} q_{i}(n) \cdot o_i \prod_{m\in \Nopenin{i}{n}} (1-o_m) ~. 
\end{equation}

\subsection{Algorithm for Finding Hit Probabilities}
Next, our goal is to 
% define
propose
an algorithm to compute hit probabilities. To this aim, we solve the following set of equations:
\begin{align}
\label{e:lambda-entry-independence}
    &\vec{\lambda^{e}} = f^{e}(\vec{o})  ~,\\ 
    % \nonumber
    % &\lambda_n^{e}= \lambda_n \left(  \prod_{i \in \Nopen{n} } (1-\newoccupancy_i) + \sum_{i\in \Nopen{n}} (1-q_n(i)) \newoccupancy_{i} \prod_{j\in \Nopenin{i}{n}} (1-\newoccupancy_{j}) \right) ~,\\
    \label{e:lambda-refreshement-independence}
    &\vec{\lambda^{r}}  = f^{r}(\vec{o}) ~,\\
    % \;\lambda_n^{r} =  \sum_{i\in \Nclosed{n}} \lambda_{i}  \prod_{m \in \Nclosedin{n}{i}} \left(1-\newoccupancy_{m}\right) ~,\\ 
        \label{e:vec-occupancy-system}
    &\vec{\occupancy} = f^{o}(\vec{\lambda}^r,\vec{\lambda}^e,t_C),\\
    %= f^{o} (\vec{o}, t_C): \; \sum_{n \in I} \newoccupancy_{n} =C ~,  \\
    \label{e:capacity_constraint}
    &\sum_{n \in I} \newoccupancy_{n} =C,\\
    \label{e:vech}
    &\vec{h} = f^{h}(\vec{o}) ~. 
    \end{align}
Equation~\eqref{e:lambda-entry-independence} follows from~\eqref{e:lambda-entry-def}, \eqref{pnei-independence}, and~\eqref{e:pne-independence}, and computes the vector of insertion rates for all items.  Equation~\eqref{e:lambda-refreshement-independence} follows from~\eqref{e:lambda-refreshement-def} and~\eqref{e:pnr-independence} and computes the vector of refresh rates.
%, i.e., the rate at which item $n$ is used to serve users requests while it is stored in cache. 
Equation~\eqref{e:vec-occupancy-system} is the vector form of~\eqref{e:occupancy-simlru-che}:
given $\vec{\lambda}^r$ and $\vec{\lambda}^e$, and the characteristic time $t_C$, it computes all occupancies. Equation~\eqref{e:capacity_constraint} expresses the capacity constraint.
Combining \eqref{e:lambda-entry-independence}--\eqref{e:capacity_constraint}, we obtain a system of $3N+1$ equations in $3N+1$ unknowns, from which we can obtain in particular the occupancies and the characteristic time $t_C$. 
Finally, once the occupancies are known Equation~\eqref{e:vech} computes the vector of hit rates according to \eqref{e:hit-rate-rnd-lru-independence}.

%and cache capacity $C$, \eqref{e:vec-occupancy-system} is used to characterize, for each item, the probability that it is present in cache; it is obtained by adding $o_n$ to both sides of the equality in \eqref{e:occupancy-simlru-che}. Finally, \eqref{e:vech} expresses the hit rate for each item as in \eqref{e:hit-rate-rnd-lru-independence}. 

%Using \eqref{e:pne-independence} in \eqref{e:lambda-entry-def} yields $\lambda_{n}^{e}$, whereas $\lambda_{n}^{r}$ is obtained from \eqref{e:lambda-refreshement-def} and \eqref{e:pnr-independence}. Equation \eqref{e:occupancy-simlru-che} along with \eqref{e:cache-capacity-constraint} become a system of $N+1$ equations in $N+1$ unknowns: the $N$ occupancies $\Vecnot{\newoccupancy_{n}}{n\in I}$ and the characteristic time $t_C$. 

% Finally, given $\vec{\lambda}^r$ and $\vec{\lambda}^e$, and cache capacity $C$, 
% function $ f^{o}(\vec{\lambda}^r,\vec{\lambda}^e,t_C)$ in~\eqref{e:vec-occupancy-system} is used to characterize, for each item, the probability that it is present in cache. Vector $\vec{\occupancy}$ is the vector of item occupancies. In this paper, we use Che's approximation to instantiate  $f^{o}(\vec{\lambda}^r,\vec{\lambda}^e,t_C)$. 

% expresses the rate at which requests for item $n$ cause that item to be inserted into the cache. Insertions occur when $n$ is requested and  none of its neighbors are found in the cache. The probability of the latter event   is given by $p_n^e$ and corresponds to the product in~\eqref{e:lambda-entry-independence}. Equation~\eqref{e:lambda-refreshement-independence} 

\begin{algorithm}[tb]
\footnotesize
\caption{Fixed point method}\label{alg:cap}
\begin{algorithmic}[1]
 \renewcommand{\algorithmicrequire}{\textbf{Input:}}
 \renewcommand{\algorithmicensure}{\textbf{Output:}}
 \REQUIRE $C$,  $\vec{\lambda},$ $\mathrm{dis}(.,.),$ $d,$  $(q_{n}(i))_{(n,i)\in I^{2}}$, stopping condition 
 \ENSURE  Estimation $\vec{o}, \vec{h} , t_C$ 
 \\ 
 \textit{Initialization}: 
\STATE Obtain $t_{C}(0)$ such that $  \sum_{n\in I}\left(1-\e^{- \lambda_{n}\cdot t_{C}(0)} \right) = C$ \label{line:ini0}
\STATE $\vec{o}(0) \gets 1-\e^{- \vec{\lambda}\cdot t_C(0)}$ 
\label{line:ini1}
\STATE $\vec{h}(0) \gets f^{h} (\vec{o}(0))$  
%  \IF {\emph{Adjust workload}} 
%   \STATE $\tilde \lambda_n \gets \sum_{i\in \Nclosed{n}} \lambda_i ,\; \forall n  \in I $ 
%  \ELSE
%   \STATE $\tilde \lambda_n \gets \lambda_n, \; \forall n \in I$
%   \ENDIF 
%      \label{line:ini0} \STATE Obtain $t_C(0)$ such that $  \sum_{n\in I}\left(1-\e^{-\tilde \lambda_{n}\cdot t_C(0)} \right) = C$
% % \FOR{$n\in I$} 
% \STATE $\newoccupancy_{n}^{0} \gets 1-\e^{-\tilde \lambda_{n}\cdot  
% \label{line:hitini}  t_C(0)}, \; \forall n \in I$ \\
% \IF {\emph{Adjust hit probabilities}} 
% \STATE $ \hit_{n}^{0} \gets 1- \prod_{i\in \Nclosed{n} } \left( 1- \newoccupancy_{n}^{0}\right), \; \forall n \in I $
% \ELSE
% \STATE $ \hit_{n}^{0} \gets \newoccupancy_{n}^{0}, \; \forall n \in I $
% \ENDIF
% \ENDFOR
\STATE $j \gets 1$
 
\WHILE{\textit{Stopping condition not satisfied}} 
 \label{line:work0}
    \STATE $\vec{\lambda^{e}}(j) \gets f^{e}(\vec{o}(j-1))$ (See \eqref{e:lambda-entry-independence})  \label{line:lambda-e}
    \STATE $\vec{\lambda^{r}}(j) \gets f^{r}(\vec{o}(j-1))$  (See \eqref{e:lambda-refreshement-independence}) \label{line:lambda-r}
\label{line:work1}
    \STATE $\textrm{Obtain } \label{line:obtaintcj1}  t_{C}(j) \textrm{ such that}:  \sum_{n\in I} (f^{o}( \vec{\lambda^{e}}(j) , \vec{\lambda^{r}}(j) , t_C(j)))_{n} =C$ (See \eqref{e:capacity_constraint},\eqref{e:vec-occupancy-system})
\label{line:obtaintcj} 
    \STATE $\vec{o}(j) \gets (f^{o}( \vec{o}(j-1), t_C(j))+\vec{o}(j-1))/2$  \label{line:occ1}  %(See \eqref{e:vec-occupancy-system})  
    \STATE  $\vec{h}(j) = f^h(\vec{o}(j))$ (See \eqref{e:vech})
    \STATE $j\gets j+1$
\ENDWHILE 
 
\RETURN $\vec{\hit}(j), \; \vec{\newoccupancy}(j), \; t_C(j)$  
\end{algorithmic} 
\end{algorithm}

To solve the system of equations~\eqref{e:lambda-entry-independence}-\eqref{e:capacity_constraint}, we rely on an iterative fixed point method (see Algorithm~\ref{alg:cap}). We begin  by guessing occupancies $\vec{o}$. In particular, we  initialize them using 
occupancies for LRU, i.e., $\vec{o}(0) = 1 - \e^{-\vec{\lambda} \cdot t_C(0)}$ where $t_C(0)$ verifies \eqref{e:cache-capacity-constraint} ($\sum_{n\in I} o_n(0) = C$)  (lines~\ref{line:ini0}-\ref{line:ini1}). 
Then, we obtain $ \vec{\lambda^{e}}(1)$  and $\vec{\lambda^{r}}(1)$ using equations \eqref{e:lambda-entry-independence} and \eqref{e:lambda-refreshement-independence}, resp. (lines \ref{line:work0}-\ref{line:work1}). Next we find  the new estimation of the occupancies $\vec{o}(1)=  f^{o}(\vec{\lambda^{e}}(j) , \vec{\lambda^{r}}(j),t_C(1))$ where $t_C(1)$ verifies $\sum_{n\in I} o_n(1) = C$ (lines \ref{line:obtaintcj}-\ref{line:occ1}). Finally, a new estimate of the vector of occupancies is computed (line \ref{line:occ1}): averaging the new prediction and the previous value is a practical trick to improve the convergence.
The same procedure is then repeated for the next iterations until a stopping condition is reached, e.g., the difference between $\vec{\occupancy}$ computed at consecutive iterations becomes smaller than a given threshold, or the maximum number of iterations is reached ($j\leq n_{\mathrm{iterations}}$).

% Equations~\eqref{e:lambda-entry-independence}-\eqref{e:vec-occupancy-system} are solved in order, i.e., each equation uses the results obtained from the previous equations. 

% In each
% iteration of Algorithm~\ref{alg:cap} we solve equations~\eqref{e:lambda-entry-independence}-\eqref{e:lambda-refreshement-independence} using the vector of hit probabilities $\vec{\occupancy}$ obtained in the previous iteration (lines~\ref{line:lambda-e}-\ref{line:lambda-r}), and then recompute $\vec{\occupancy}$ using~\eqref{e:vec-occupancy-system} (lines~\ref{line:obtaintcj}-\ref{line:occ1}). 

%Popularity Figures 
\begin{figure}[b]
    \begin{subfigure}{0.46\linewidth}
  \centering
  \includegraphics[width=0.99\linewidth]{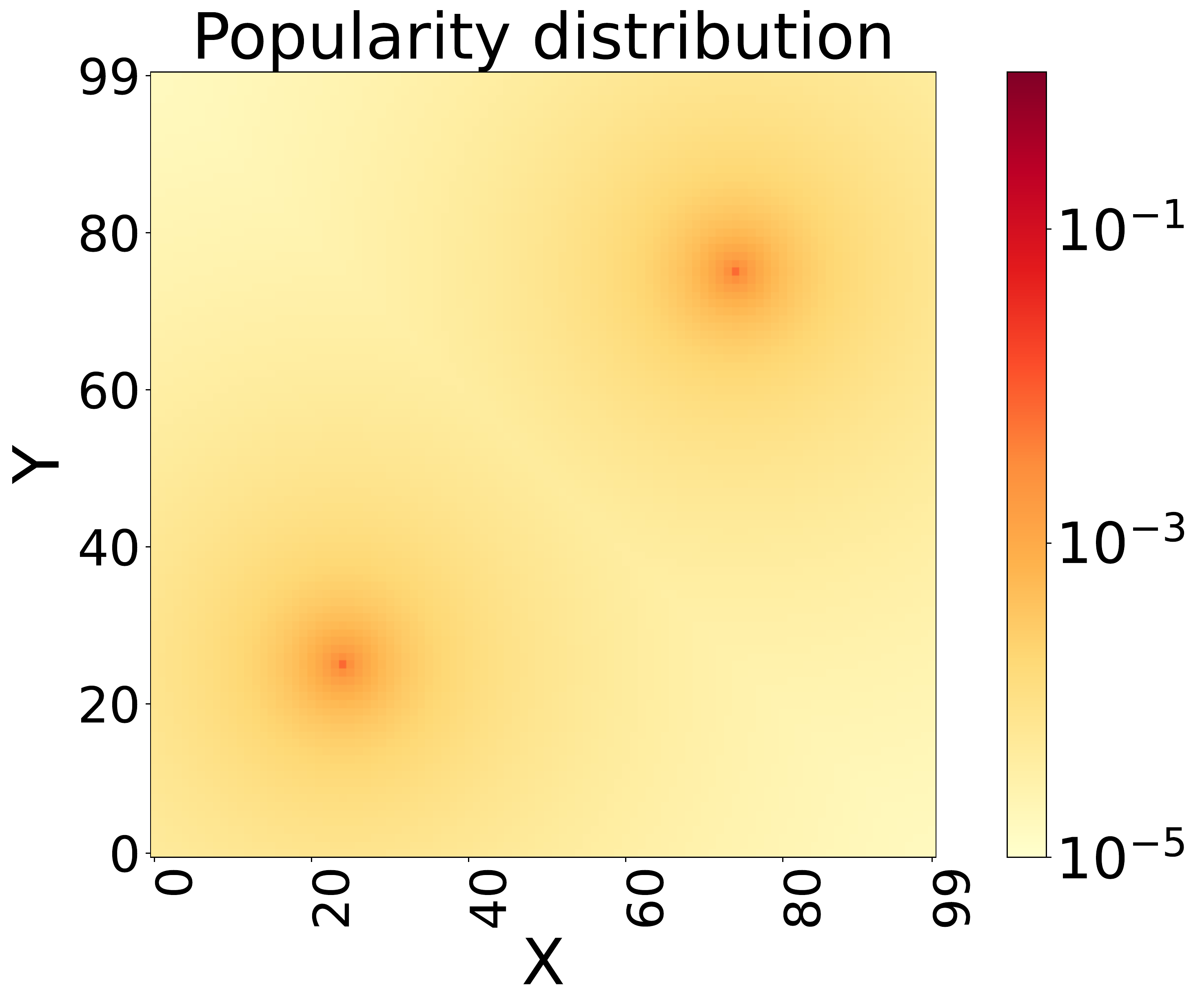}
\caption{$\alpha=1.4$}
\label{fig:popularity-alpha1.4}
\end{subfigure}
\hfill
\begin{subfigure}{0.46\linewidth}
  \centering
  \includegraphics[width=0.99\linewidth,keepaspectratio]{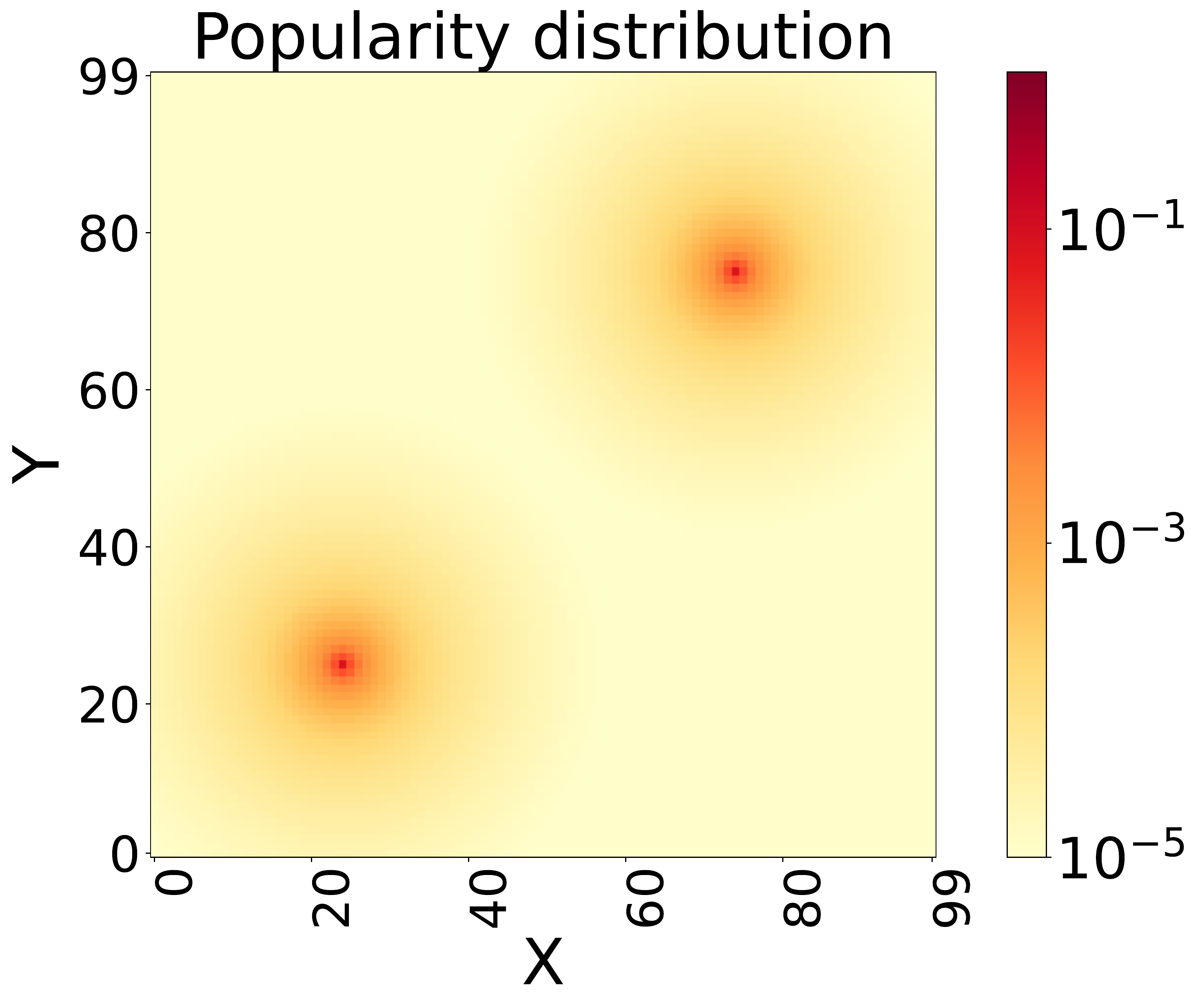}
\caption{$\alpha=2.5$}
\label{fig:popularity-alpha2.5}
\end{subfigure}
    \caption{Synthetic traces: Spatial popularity distribution.}
    \label{fig:popularities}
\end{figure}

%Hit rate synthetic and Amazon trace. 
\begin{figure*}[tb]
    \begin{subfigure}{0.32\linewidth}
  \centering
  \includegraphics[width=0.99\linewidth]{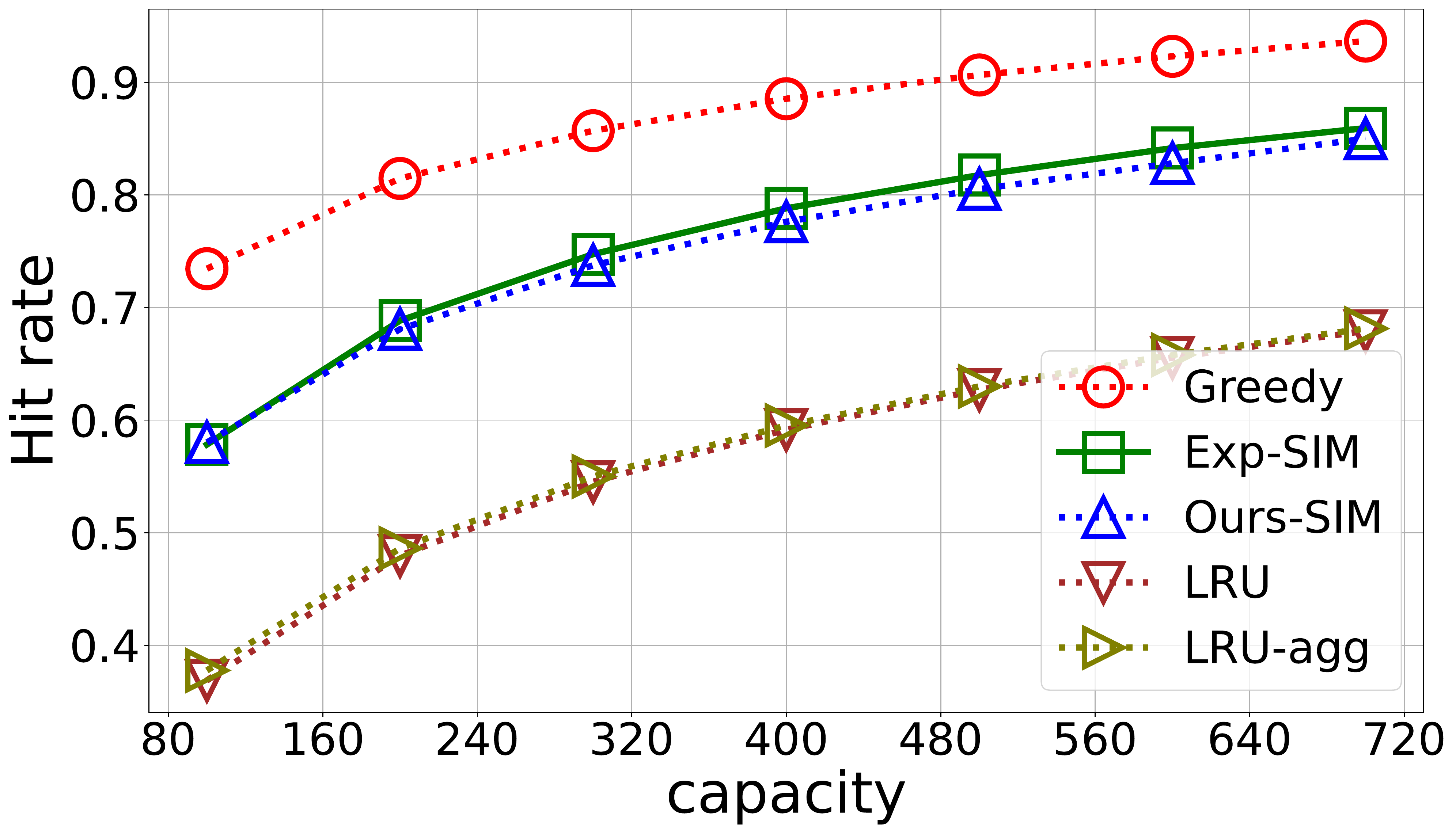}
\caption{Synthetic trace,  $\alpha = 2.5$, $d=1$, \\$25$ iterations}
\label{fig:hit-rate-d1}
\end{subfigure}
\hfill
\begin{subfigure}{0.32\linewidth}
  \centering
  \includegraphics[width=0.99\linewidth, keepaspectratio]{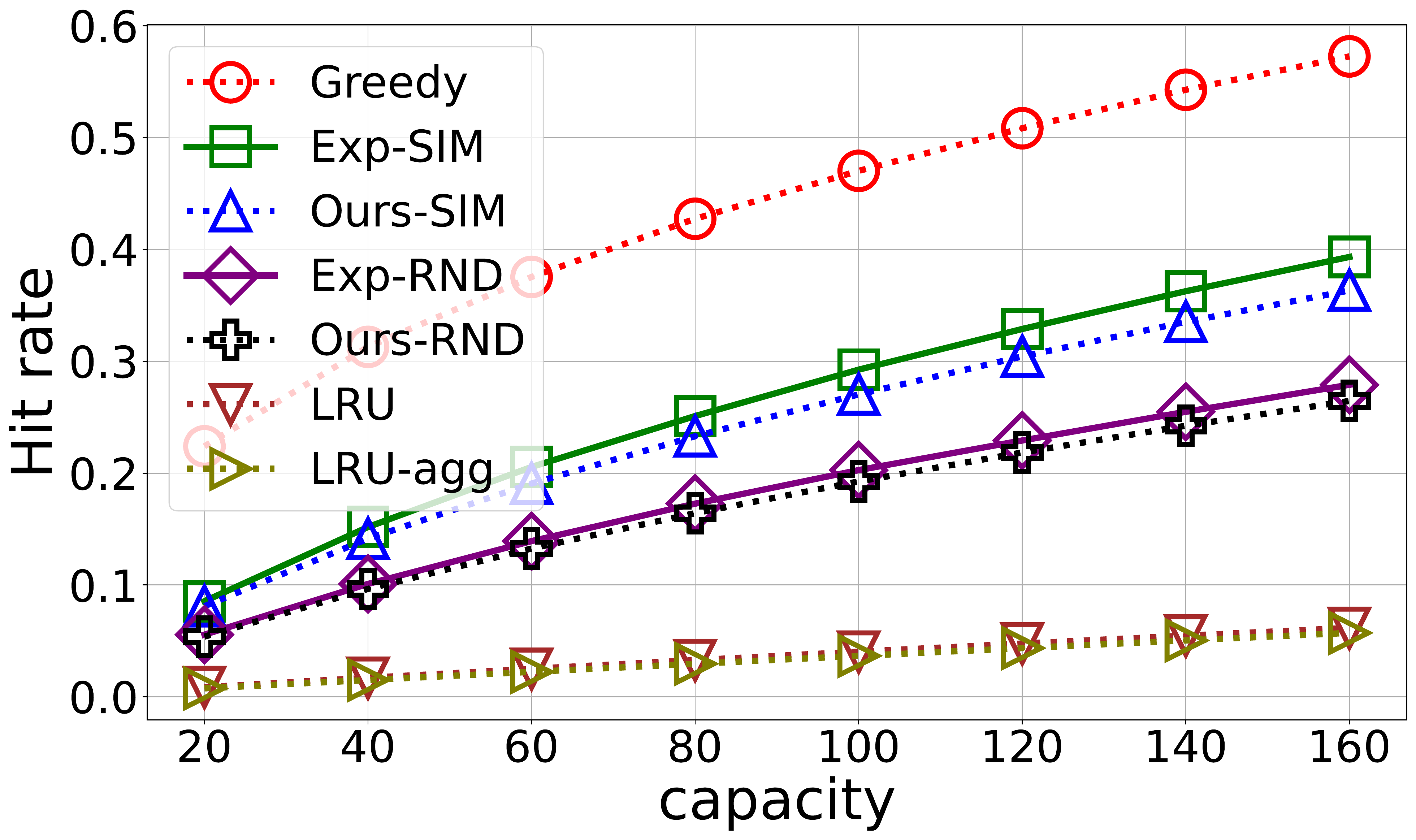}
  \caption{Synthetic trace, $\alpha=1.4$, $d=2$,\\  $15$ iterations}
  \label{fig:hit-rate-d2}
\end{subfigure}
\hfill
 \begin{subfigure}{0.32\linewidth}
      \centering  
    \includegraphics[width=0.99\linewidth,keepaspectratio]{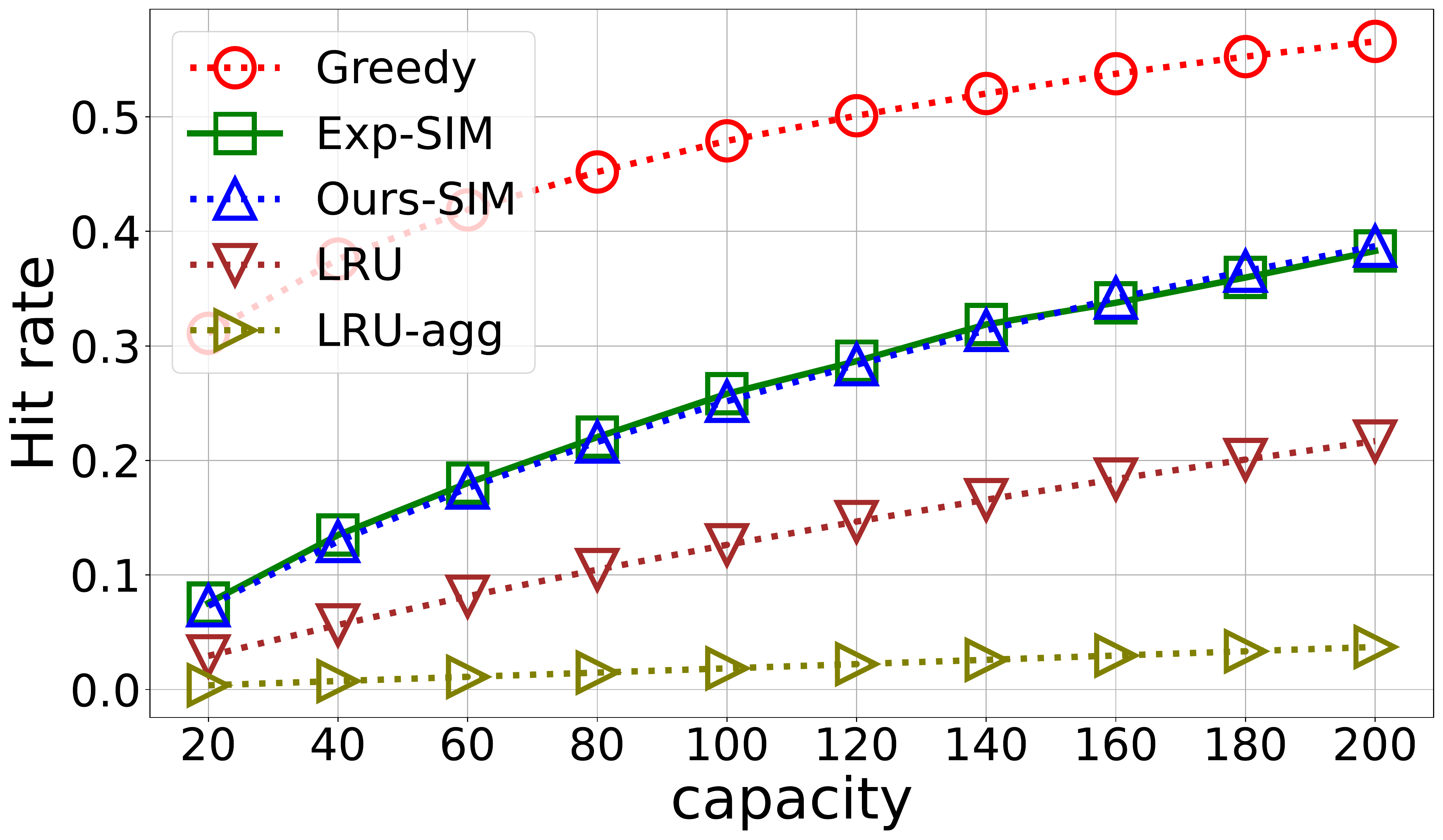}
    \caption{Amazon trace, $d=300$, $40$ iterations\\~}
    \label{fig:hit-rate-amazon}
\end{subfigure}
    \caption{Hit rate versus cache capacity.}
    \label{fig:hit-rates}
\end{figure*}

\subsection{Benchmarks and Alternative Approaches}
\label{ss:alternative solutions}
%Next, we consider alternative solutions to determine hit probabilities.

In what follows, we compare hit rate estimates provided by RND-LRU or SIM-LRU using  Algorithm~\ref{alg:cap} with the hit rate estimations for LRU and for the optimal static allocation. We also propose an alternative approach to estimate RND-LRU's hit rate.

\textbf{LRU.} 
%We consider a classical  LRU mechanism, ignoring the fact that 
% alternative items may be used to serve a request.
%items may be used to serve their neighbours' requests. 
The hit rate and the occupancy for an item $n$ are computed using Eq.~\eqref{e:hit-rate-item-lru-che} and $t_C$ is deduced using the cache capacity constraint  given by Eq.~\eqref{e:cache-capacity-constraint}. 
% This corresponds to
% \begin{align}
%     \Nopen{n} &= \emptyset, \quad \Nclosed{n}= \{n\}
%     % , \quad p_n^e= 1, \quad p_n^r(i) = 0, n\neq i.  
% \end{align}
% Note that in this case~\eqref{e:hit-rate-item-lru-che} and~\eqref{e:occupancy-simlru-che}  are identical, as $\lambda_n^e=\lambda_n^r=\lambda_n.$ Therefore, Algorithm~\ref{alg:cap} executes  lines~\ref{line:ini00}-\ref{line:ini1}, with \emph{Adjust workload} and \emph{Adjust hit probability} set to \emph{False}, noting that line~\ref{line:occ1} is equivalent to line~\ref{line:hitini}.

\textbf{Optimal Static Allocation.} 
The maximum hit rate obtainable by a static allocation under similarity caching can be obtained solving a maximum weighted coverage problem. We consider, as in SIM-LRU, that each item can be used to satisfy any request for items closer than $d$.
The maximum weighted coverage problem takes as input a capacity $C$, a set of items $I$, with $N=|I|$, their corresponding weights $W=(w_i)_{i\in I}$ and a set of sets $R=\{R_1,\ldots, R_N \}$ such that $R_i \subset I$.  The objective is to find a set $\sigma^{*}\subset \{ 1,\ldots , N\} $  such that: $ \sigma^{*} = \argmax_{\sigma \subset \{ 1,\ldots , N\}: |\sigma|\leq C  } \sum_{i\in \cup_{j\in \sigma } R_j}  w_i$.
% \begin{equation}
%     \sigma^{*} = \argmax_{\sigma \subset \{ 1,\ldots , n\}: |\sigma|\leq C  } \sum_{i\in \cup_{j\in \sigma } R_j}  w_i ~.
% \end{equation}
Finding the best static allocation is equivalent to solving a maximum weighted coverage problem, with   
weights $w_i= \lambda_i$ for $i\in I$, $C$ the cache capacity, and $R$ the set of neighbours for each item, i.e., $R= \{\Nclosed{n} \}_{n\in I}$. 
%SIM-LRU hit rate is an upper bound also for  then solving the maximum weighted coverage problem is equivalent to finding the optimal static similarity cache allocation in terms of the hit rate. Under the IRM model, storing the optimal static allocation is a better strategy than SIM-LRU in terms of the hit rate. 
The maximum weighted coverage problem is known to be NP-hard. In practice, a popular greedy algorithm guarantees a $(1- 1/e)$ approximation ratio~\cite{nemhauser1978analysis}.
%approximation algorithm provides a $\sigma'$ that is guaranteed to have a hit rate larger than $(1-\mathrm{e}^{-1})H^{*}$, where $H^{*}$ is the hit rate achieved by the optimal static allocation \cite{nemhauser1978analysis}. 

The greedy algorithm chooses at the first step the set with the largest coverage $c_m=\max_{n\in I} \sum_{i \in R_n^{0}=R_{n}} p_i$. If $R_{o}^{0}$ is the set chosen at the first iteration, at  the next iteration all the sets are updated in such a away that they do not contain any item in the set $R_{o}^{0}$, i.e. $R_n^{1} = R_{n}^{0} \setminus R_{o}^{0}$. The same procedure is repeated until $C$ sets are collected or all the sets are chosen.

% This algorithm is described in \cite{nemhauser1978analysis} and its a $1-\frac{1}{e}$ approximation algorithm for the maximum weighted coverage problem.   
% We consider a static item allocation to lower bound the hit probability. 
%   Let $c_m$ be the coverage of  item $m$,   defined as the sum of the popularities of items can be  used to serve  a request to $m$, i.e., $c_m=\sum_{i \in \mathcal{N}[m]} p_i.$ 
% The static allocation is obtained in a greedy fashion, where items with largest coverage  $c_m$ are prioritized for insertion.

% \textbf{LRU experimental. (LRU exp)} This is similar to the previous solution. However, instead of relying on the characteristic time to approximate LRU, we use simulations for that purpose. Our purpose here is to assess conditions under which the characteristic time can or cannot approximate LRU behavior.

%\textbf{Naive application of characteristic time approximation.}
\textbf{LRU with aggregate requests.}
% This corresponds to running one  iteration of Algorithm~\ref{alg:cap}, i.e., setting as stop condition $j > 1$, and skipping line~\ref{line:obtaintcj} so as to use the characteristic time $t_C$ obtained using classical LRU to approximate similarity cache behavior.  Our purpose here is to assess the impact of re-estimating the characteristic time, accounting for item similarity,  before assessing hit probabilities. %
%
% This corresponds to running lines~\ref{line:ini00}-\ref{line:ini1}, with \emph{Adjust workload} set to \emph{True} and \emph{Adjust hit probability} set to \emph{False}: 
Under SIM-LRU  an item is refreshed by the requests for all its neighbours. A naive approach to study a SIM-LRU cache is then to consider that it operates as a LRU cache with equivalent request rates for each item equal to the sum of the request rates for all items in its neighborhood. One can then use the TTL approximation for LRU, leading to the following formulas:
%'s hit rate is to assume that the similarity cache operates as a classic LRU cache, where 
%This approach computes the hit rate and the occupancy in a similar way to the TTL approximation for LRU while accounting for the approximate hits thanks to an item's neighbours. More specifically, the hit rate and the occupancy are computed as:  
    \begin{equation}\label{e:hit-rate-naive-che}
        \hit_{n} = 1- \e^{-\sum_{i\in \Nclosed{n}} \lambda_i \cdot t_C } , \quad \newoccupancy_{n} =\hit_{n}~.
    \end{equation}

\section{Numerical Evaluation}
\label{s:experiments}

We evaluate the efficiency of the proposed fix point method (Algorithm \ref{alg:cap}) to predict the hit rate on synthetic traces and on an Amazon trace \cite{sabnis2021grades}.
%The synthetic traces description (Topological space  stream of requests)
For the synthetic traces, each item corresponds to two features, characterized by a point in a grid, $I=[0..99]^{2}$ (e.g. Fig.~\ref{fig:popularities}). The total number of items is $|I|= 10^{4}$, and the dissimilarity function between items  $\mathrm{dis}(\cdot,\cdot)$ is the Euclidean distance. Neighbours of item $(x,y)$ at the same distance are ordered counterclockwise starting from the item to the right, i.e., from $(x+a,y)$ with $a>0$.
%In case of a tie, i.e. $\mathrm{dis}(i,j_1) =\mathrm{dis}(i,j_2)$, items are ordered  an angle determining the position of points $j_1$ and $j_2$ in the circle centred on $i$ is used to determine whether it is $j_1$ or $j_2$ who is going to serve $i$'s requests when both $j_1$ and $j_2$ are in the cache and $i$ is not in the cache. 
Note that for   similarity thresholds $d\in \{ 1,2\}$ the proposed distance produces an ordering  equivalent to  Manhattan distance (MD), with  MD ties  broken in such a way that  items  in same row or column   have higher distance than their counterparts.

We generate a stream of $r$ requests for items in $I$ in an IRM fashion \cite{irm-fagin-1977},   $r=2\cdot 10^{5}$. The popularity distribution for an item $n=(x,y)$ is given by 
\begin{equation}
\label{e:arrival-rates-square-two-hot-regions}
     p_{(x,y)} \sim  \left(\min \left \{\mathrm{dis}(n,(24,24)), \mathrm{dis}(n,(74,74)) \right\} +1\right)^{-\alpha} ~,
\end{equation}
where %$n_1$ and $n_2$ are two items in the grid with coordinates $(24,24)$ and $(74,74)$ and 
$\alpha$ is a parameter controlling the skew of the popularity distribution. 
%and $n_1$ and $n_2$ are items  with coordinates $(24,24)$ and $(74,74)$.  
%We generate streams for values of $\alpha \in \{1.4, 2.5 \} $.  
%that is shown for $\alpha \in \{1.4, 2.5 \} $  in 
Fig.~\ref{fig:popularities} illustrates the cases $\alpha \in \{1.4, 2.5 \}$.  
% The Amazon trace description ((Topological space  stream of requests))

For the Amazon trace,~\cite{mcauley2015image} proposes a scheme to embed the images of Amazon products in a $100$-dimensional
space, where the Euclidean distance reflects  dissimilarity between two items. Then,~\cite{sabnis2021grades} reports the number of reviews per product, and equates it to   product request rates. Inspired by this methodology, we   leverage the empirical request  probabilities, and use it to generate a corresponding   IRM stream of requests. 

%EXP computation 

Given the workloads, we  evaluate  similarity cache mechanisms employing SIM-LRU with threshold similarity $d\in \{ 1,2\}$ for the synthetic traces and $d=300$ for the Amazon trace. For the synthetic trace with $d=2$, we also evaluate RND-LRU where the probabilities $q_n(i)$ are mapped to $\mathrm{dis}(n,i)$ as $\left( \left[ 1,\frac{1}{2},\frac{1}{4}\right], \left[ 1, \sqrt{2}, 2 \right] \right)$.
% \begin{equation}
%  q_n(i) = 
%     \begin{cases}
%   1 \textit{ if } \mathrm{dis}(n,i)= 1 \\ 
%   \frac{1}{2} \textit{ if } \mathrm{dis}(n,i)= \sqrt{2} \\ 
%   \frac{1}{4} \textit{ if } \mathrm{dis}(n,i)= 2~. 
%     \end{cases}
% \end{equation}
% such that the neighbours for an item $n$ are $\Nclosed{n}\triangleq \{i\in I: \mathrm{dis}(n,i)<d\}$,
The $95\%$ confidence intervals were smaller than $1.2 \cdot 10^{-3}$ in all the considered synthetic experiments for the hit rate computation. In all experiments, we refer to the empirical hit rates for SIM-LRU and RND-LRU as `Exp-SIM' and `Exp-RND', respectively.

%Theory-computation 
For all the theoretical computations of the hit rate, the arrival rates $\vec{\lambda}$ for items are taken equal to the corresponding request  probabilities. Our approach uses Algorithm \ref{alg:cap} to compute the hit rates for each item, $\vec{h}$, and then deduces the cache hit rate $H$. We refer to the latter estimate, for SIM-LRU and RND-LRU, as `Ours-SIM' and `Ours-RND', respectively. Alternative methods that could be used to estimate the hit rate are presented in Sec.~\ref{ss:alternative solutions}. We refer to the TTL approximation for LRU as `LRU', %\cite{fricker2012versatile,che2002hierarchical}, 
LRU with aggregate requests as `LRU-agg', and the greedy algorithm as `Greedy'. 
% The approach that considers the aggregation of rates %takes into account similarity and uses TTL approximation 
% is denoted as  and the greedy algorithm as `Greedy'. 
The numerical values used for all the experiments are summarized in Table~\ref{tab:parameters-experiments}.           

In Fig.~\ref{fig:hit-rates}, we show the empirical hit rate along with its predictions, including those predictions obtained with our approach, for the two synthetic settings and for the Amazon trace. In  the considered settings,  `Greedy' overestimates the hit rate.  `LRU' and `LRU-agg', in contrast,  underestimate it. 

`Ours-SIM' and `Ours-RND'  clearly outperform all the  alternative approaches presented in Sec.~\ref{ss:alternative solutions} in estimating the empirical hit rate, while tending to  underestimate it.
`LRU' does not take into account the similarity between items, hence the gap between `LRU' and `Exp-SIM' shows us the benefits of similarity caching over exact caching.
For the synthetic settings in Figs.~\ref{fig:hit-rate-d1} and~\ref{fig:hit-rate-d2}, `LRU' and `LRU-agg' achieve similar hit rates. This is possibly due to the choice of the popularity distribution (see \eqref{e:arrival-rates-square-two-hot-regions}) where a popular item $n$ and its neighbours have similar popularities: $\tilde \lambda_{n}=\sum_{i\in \Nclosed{n}} \lambda_i \approx |\Nclosed{n}| \lambda_n$, implying that  $\vec{ \tilde \lambda} \approx f(d) \vec{\lambda}$, which corresponds to the case wherein it is equivalent to  computate $h_n$ using either `LRU' or `LRU-agg'. 
% and thus also the global hit rate.   

% `LRU-agg' considers that if two items are neighbours then they have the same occupancy whereas     

% this can be expected as this algorithm works well in practice in predicting the optimal static allocation which is an upper bound on the hit rate for an IRM process. 
% `LRU' and `LRU-agg' seems to underestimate the hit rate. 

% `Ours' in the other hand seems to be predicting well the empirical hit rate with comparison to other approaches. It seems though that our approach provides a lower bound this might be interpreted by looking at the occupancies. 

%\vspace{4mm} 

\begin{table}[t]
   \caption{Parameters of the experiments}
   \label{tab:parameters-experiments}
   \centering
\begin{tabular}{ |l|l|l| } 
 \hline
 Variable & Synthetic traces  & Amazon trace \\
 \hline
 \hline
 $I$ & $[0..99]^{2}$ & Products  \\
 $N=|I|$ & $10^{4}$ & $\approx 10^{4}$  \\
 $\lambda_n$ & \eqref{e:arrival-rates-square-two-hot-regions} & Empirical \\
 $\mathrm{dis}(\cdot, \cdot)$ &Euclidean distance & Euclidean distance \\
 $d$&  $1$ and $2$ & $300$ \\
 Number of requests $r$ & $2\cdot 10^{5}$  & $\approx 10^{5}$ \\ 
 $95\%$ confidence intervals & $\approx 10^{-3}$ & ---  \\ 
 Number of iterations & $25$ and $15$  & $40$\\ 
 \hline
\end{tabular}  
\end{table}

%Occupancies Figures 
\begin{figure}[t]
    \begin{subfigure}{0.48\linewidth}
  \centering
  \includegraphics[width=0.99\linewidth]{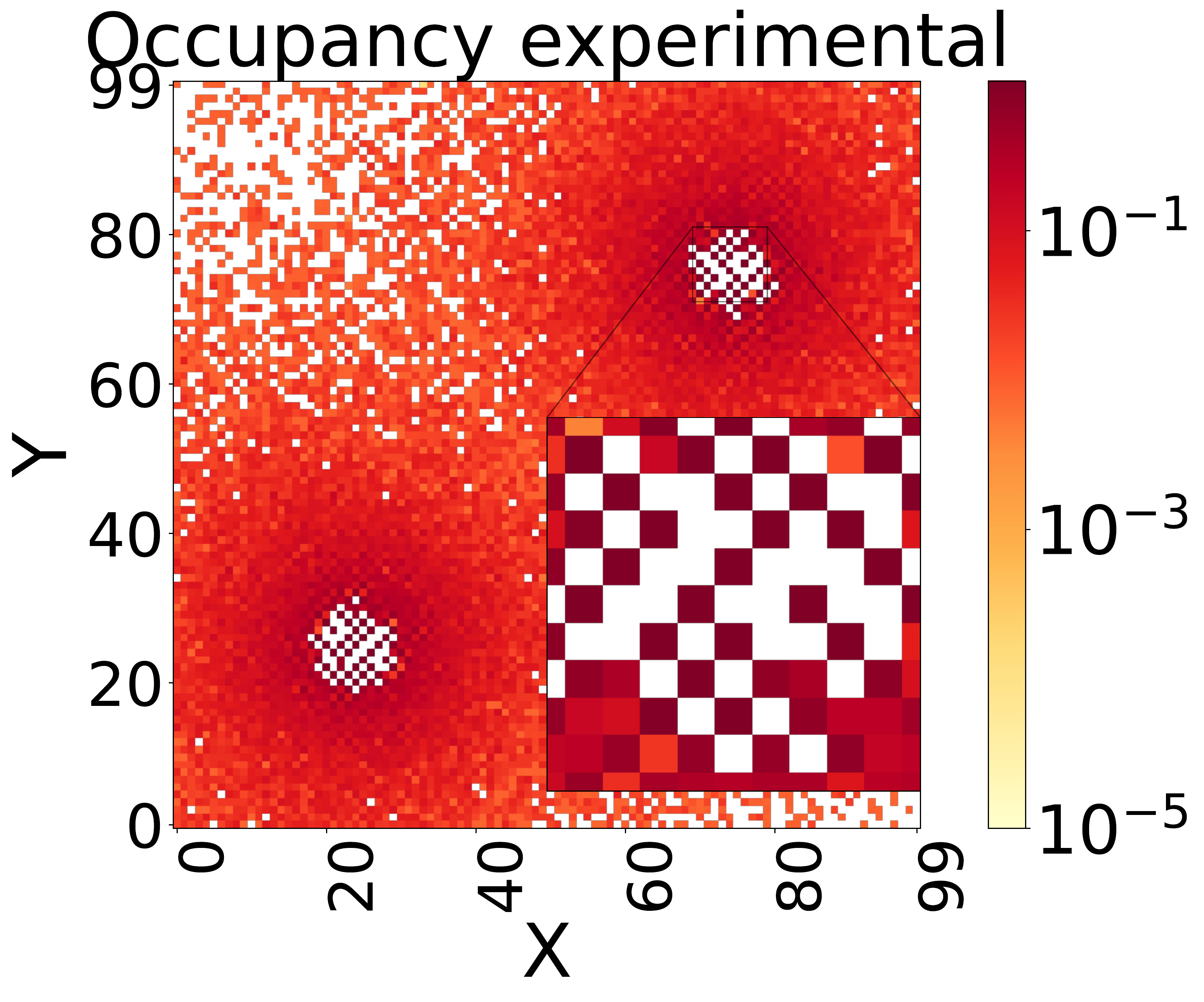}
\caption{$r=2\cdot 10^{5}$}
\label{fig:occupancy-exp-c500-d1}
\end{subfigure}
\hfill
\begin{subfigure}{0.48\linewidth}
  \centering
  \includegraphics[width=0.99\linewidth,keepaspectratio]{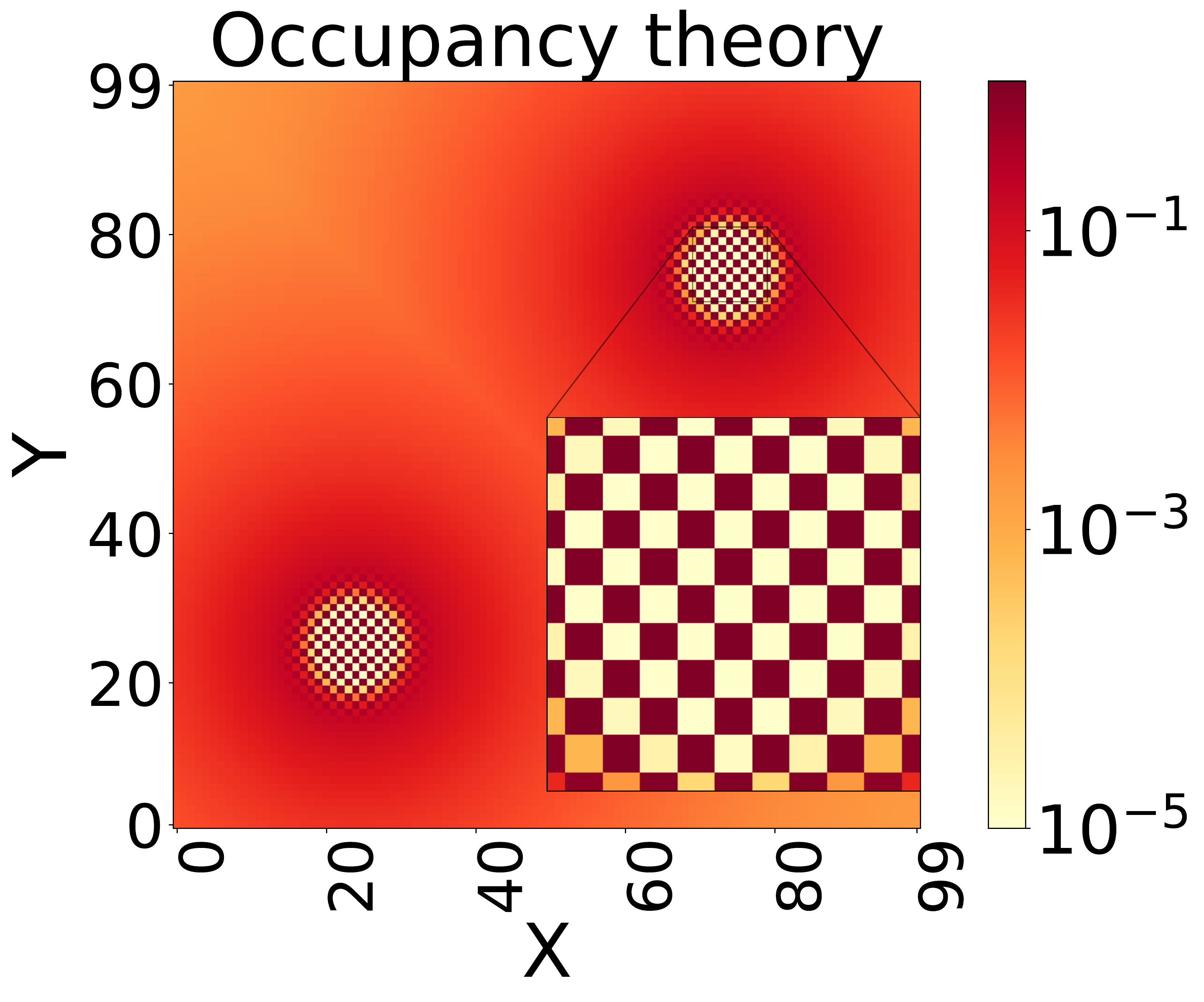}
\caption{$25$ iterations}
\label{fig:occupancy-theory-c500-d1}
\end{subfigure}
    \caption{Synthetic trace occupancies: $C=500$, $d=1$, $\alpha=2.5$.}
    \label{fig:occupancies}
\end{figure}  

%Refresh and entry rates 

% \begin{figure}[tb]
%     \begin{subfigure}{0.46\linewidth}
%   \centering
%   \includegraphics[width=0.99\linewidth]{Images/refresh_rate_n=10000,a=2.5,c=500,k=100,d=1.0,s=-2,b=1,n_iter25,f_momentum=True,.jpg}
% \caption{Refresh rate}
% \label{fig:refresh-rate-c500-d1}
% \end{subfigure}
% \hfill
% \begin{subfigure}{0.46\linewidth}
%   \centering
%   \includegraphics[width=0.99\linewidth,keepaspectratio]{Images/entry_rate_n=10000,a=2.5,c=500,k=100,d=1.0,s=-2,b=1,n_iter25,f_momentum=True,.jpg}
% \caption{Entry rate}
% \label{fig:entry-rate-c500-d1}
% \end{subfigure}
%     \caption{Entry and refresh rates: $C=500,\; d=1 ,\; \alpha=2.5$}
%     \label{fig:refresh-entry-rates}
% \end{figure}

%Characteristic time $t_C$ in different iterations. 

\begin{figure}[tb]

\begin{subfigure}{0.49\linewidth}
  \centering
  \includegraphics[width=0.99\linewidth]{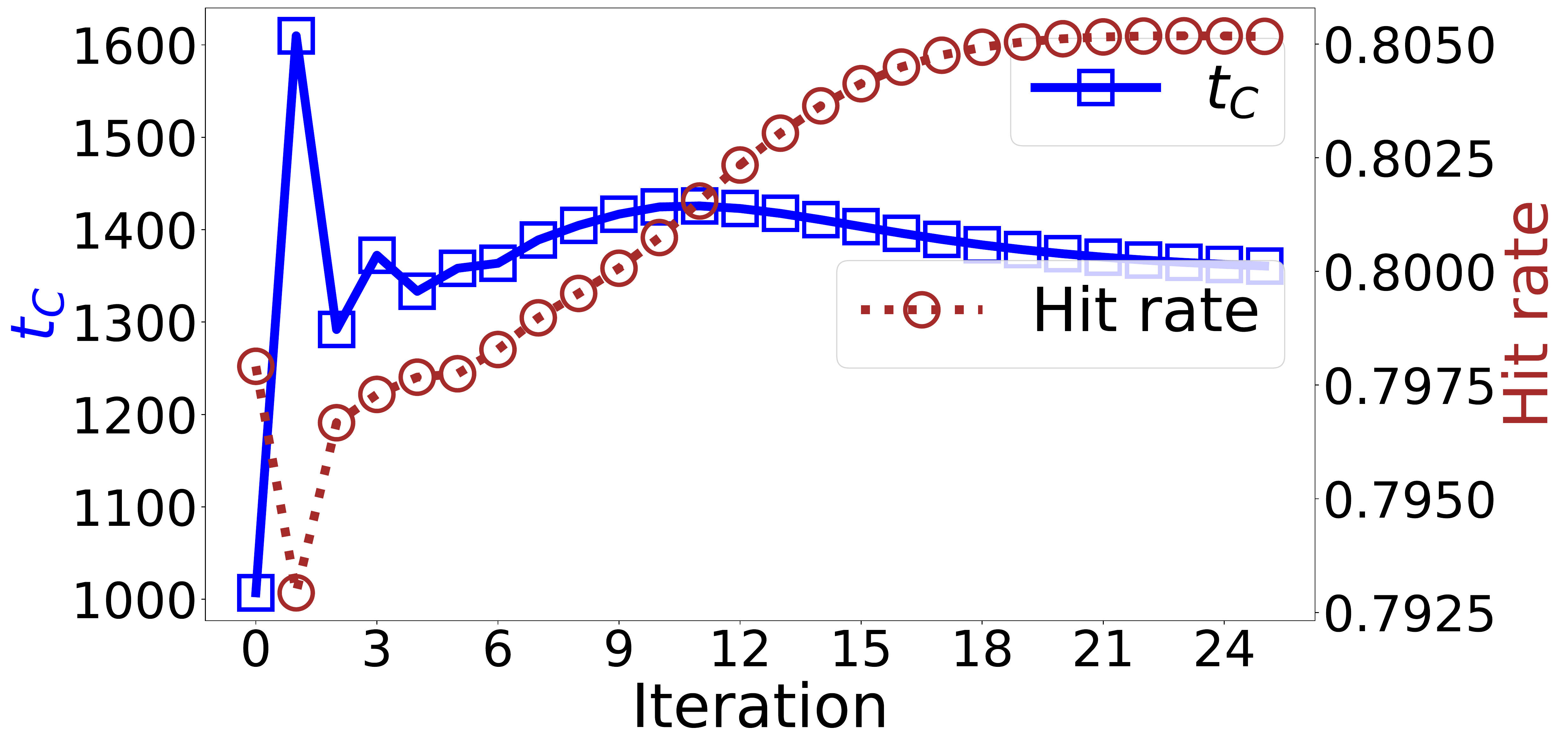}  
\caption{Synthetic trace, $ \alpha = 2.5$, $d=1$, $C=500$}
\label{fig:t_C-C500-d1} 
\end{subfigure}
% \hfill 
%     \begin{subfigure}{0.32\linewidth}
%   \centering
%   \includegraphics[width=0.99\linewidth]{Images/tc_n=10000,a=1.4,c=100,k=100,d=2.0,s=-2,b=1,n_iter15,f_momentum=True,.jpg}
% \caption{Synthetic trace, $ \alpha = 1.4$,  $d=2$, $C=100$}
% \label{fig:tc-c100-d2} 
% \end{subfigure}
\hfill 
 \begin{subfigure}{0.49\linewidth}
     \centering
    \includegraphics[width=0.99\linewidth]{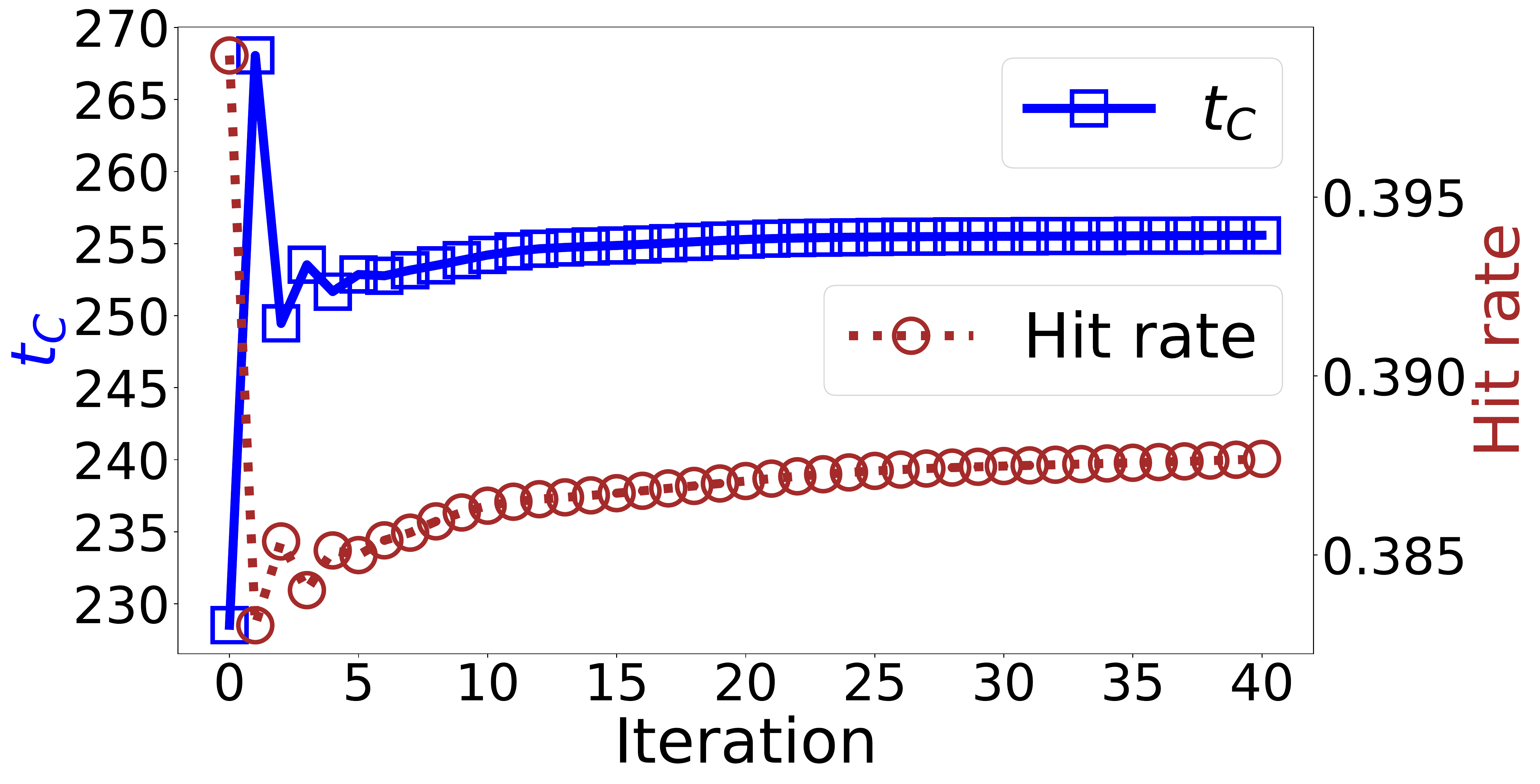}
    \caption{Amazon trace, $d=300$,  $C=200$ } 
    \label{fig:tc-amazon}
    \end{subfigure}
    \caption{Characteristic time $t_C$ and hit rate in different iterations of Algorithm \ref{alg:cap} for SIM-LRU.}
    \label{fig:characteristic time t_C}
\end{figure}

To shed further insight on why our approach underestimates the hit rate,  Fig.~\ref{fig:occupancies} shows the empirically estimated occupancy vector   and the one produced by Algorithm~\ref{alg:cap}.
The proposed algorithm broadly captures the empirical occupancy  patterns, but  with  subtleties regarding symmetries. In particular, the zoom on Fig.~\ref{fig:occupancy-theory-c500-d1} shows that our approach produces a regular chess board  pattern. Some items are predicted to stay almost all the time in the cache while their $4$ neighbours are predicted to spend virtually no time in it. 
% (if $o_n\approx 1 \implies o_m = 0 \forall m\in \Nopen{n}$)
The corresponding empirical occupancy on Fig.~\ref{fig:occupancy-exp-c500-d1} shows a less symmetric pattern, implying that in this setup SIM-LRU is able to satisfy a group of   requests using a smaller number of cache slots when compared against what is predicted by our approach. This, in turn, partially explains why our approach underestimates the hit~rate.

 Fig.~\ref{fig:characteristic time t_C} shows the evolution of   characteristic time $t_C$ and   hit rate $H$ over different iterations. We observe that   estimates of $H$ and $t_C$ by our algorithm converge  in few iterations (less than $50$), under all  the considered scenarios. Note that $t_C(0)$, the value of $t_C$ at iteration $0$,  is also the value of $t_C$ for `LRU' (see Eqs.~\eqref{e:hit-rate-item-lru-che} and~\eqref{e:cache-capacity-constraint}). In addition, across all experiments,  $t_C$ for SIM-LRU using Algorithm \ref{alg:cap} converges to a  value larger than $t_C(0)$.
Indeed, under `LRU', $t_C$ is bounded by the  time required for $C$ distinct items to be requested. For SIM-LRU and RND-LRU, in contrast,    after  $C$ distinct items are requested, an item previously in cache  can  remain there, despite not serving any requests. This occurs due to approximate hits,    explaining why    $t_C$ is larger   for SIM-LRU than   `LRU'.

\section{Conclusion}
\label{Conclusion}

We proposed the first algorithm to estimate the hit rate for popular and simple dynamic policies for similarity caching: SIM-LRU and RND-LRU, under the IRM model. Our experimental benchmark shows that our approach outperforms simple methods one can think of to predict the hit rate. Our approach builds on solving a system of equations using a fixed point method. % but we do not prove the convergence of our algorithm nor provide the conditions under which convergence is achieved. 
Although our algorithm converged in our experiments, the study of the conditions for convergence is deferred for  future work.
%Moreover, our model does not take into account 
%Acknowledging that,
In addition, note that when using SIM-LRU or RND-LRU two items whose dissimilarity is smaller than $d$ can not be simultaneously cached. %at the same time. In future work
We envision to modify our algorithm to take this fact into account. %that two neighbours can not cohabit the cache at the same time. 
Furthermore, we aim to %characterize the convergence of our algorithm and also 
investigate  the asymptotics of the  TTL approximation error, similarly to what was done in \cite{fricker2012versatile,jiang18} for classical caches.

% which might explain why our algorithm seems to underestimate the hit rate.      

% \begin{itemize}
%     \item The proposal of an algorithm to estimate the hit rate of SIM-LRU and RND-LRU popular and simple schemes for similarity caching policies.  
    
%     \item Our algorithm provides precise estimation of the hit rate and outperforms naive approaches to estimate the hit rate. 
    
%     \item We do not provide proof for the convergence of our algorithm or conditions for convergence. 
    
%     \item We assume that the events $\{i\in S \}_{i\in \Nclosed{n}}$ are independent, which is not true when two of your neighbours $i,j$ are neighbours of each other ($i\in \Nopen{j}$) and in this case clearly the events $\{ i\in S \}$ and $\{ j\in S\}$ are disjoint. 
    
%     \item Exploring the validity of the Che approximation for similarity caching similarly to what is done for LRU in \cite{fricker2012versatile}
% \end{itemize}

\section*{Acknowledgement}
 This project was financed in part by CAPES,  CNPq and FAPERJ Grant JCNE/E-26/203.215/2017.

\bibliography{references.bib}
\bibliographystyle{ieeetr}

\end{document}